\def\rfr#1{eq. (\ref{#1})}

\def\dert#1#2{\frac{{{d}}{#1}}{{{d}}{#2}}}              

\def\virg#1{``#1''}

\def\eqi{\begin{equation}}
\def\eqf{\end{equation}}
\def\eqia{\begin{eqnarray}}
\def\eqfa{\end{eqnarray}}
\def\rp#1#2{{#1\over#2}} \def\lb#1{\label{#1}}

\def\bds#1{\boldsymbol{#1}}

\documentclass{aastex}
\usepackage{url}
\usepackage{amsmath,amsthm,amscd,amssymb}
\usepackage{latexsym}
\usepackage{graphicx,epsfig}

\RequirePackage{color}

\newcommand{\emaila}{lorenzo.iorio@libero.it}

\begin{document}

\title{Solar system constraints on planetary Coriolis-type effects induced by rotation of distant masses}
\shortauthors{L. Iorio}

\author{Lorenzo Iorio\altaffilmark{1} }
\affil{F.R.A.S.: Viale Unit\`{a} di Italia 68, 70125, Bari (BA), Italy.}

\email{\emaila}

\begin{abstract}
We phenomenologically put local constraints on the rotation of distant masses by using the planets of the solar system. First, we analytically compute the orbital secular precessions induced on the motion of a test particle about a massive primary by a Coriolis-like force, treated as a small perturbation of first order in the rotation, in the case of a constant angular velocity vector $\bds\Psi$ directed along a generic direction in space. The semimajor axis $a$ and the eccentricity $e$ of the test particle do not secularly change, contrary to the inclination $I$, the longitude of the ascending node $\Omega$, the longitude of the pericenter $\varpi$ and the mean anomaly $\mathcal{M}$. Then, we compare our prediction for $\left\langle\dot\varpi\right\rangle$ with the corrections $\Delta\dot\varpi$ to the usual perihelion precessions of the inner planets recently estimated by fitting long data sets with different versions of the EPM ephemerides. We obtain as preliminary upper bounds $|\Psi_z|\leq 0.0006-0.013$ arcsec cty$^{-1}$, $|\Psi_x|\leq 0.1-2.7$ arcsec cty$^{-1}$, $|\Psi_y|\leq 0.3-2.3$ arcsec cty$^{-1}$. Interpreted in terms of  models of space-time involving cosmic rotation, our results are able to yield constraints on cosmological parameters like the cosmological constant $\Lambda$ and the Hubble parameter $H_0$ not too far from their values determined with cosmological observations and, in some cases, several orders of magnitude better than the constraints usually obtained so far from space-time models not involving rotation. In the case of the rotation of the solar system throughout the Galaxy, occurring clockwise about the North Galactic Pole, our results for $\Psi_z$ are in disagreement with the expected value of it at more than $3-\sigma$ level.
Modeling the Oort cloud as an Einstein-Thirring slowly rotating massive shell inducing Coriolis-type forces inside yields unphysical results for its putative rotation.
 \end{abstract}


\section{Introduction}
According to \citet{Mach}, the fictitious forces, proportional to the inertial mass $m$ of a body, which arise in a local non-inertial frame are merely due to a state of relative rotation $\bds\Psi$ among the considered frame and the distant masses of the universe, contrary to the Newtonian picture attributing them to the rotation of the frame with respect to the absolute space. If at a given point the average rotation of the rest of the universe yields a given amount of non-inertiality, in another place the situation will be, in general, different because of the local average rotation of the remote masses over there. For accounts of the history of the Machian ideas, the reader is referred to \citet{libro} and \citet{io}. Anyway, Mach neither had at disposal nor developed a mathematical theory supporting his views.

The situation changed with the advent of the Einsteinian investigations for a relativistic theory of gravitation. As \citet{Moller} writes,  Einstein advocated
a  new  interpretation  of  the  fictitious  forces  in  accelerated  systems  of  reference:
instead  of  regarding  them  as  an  expression  of  a  difference  in  principle  between  the
fundamental  equations  in  uniformly moving  and  accelerated  systems,  he  considered
both  kinds  of  reference  to  be  completely  equivalent  as  regards  the  form  of  the
fundamental  equations:  and  the  \virg{fictitious}  forces were  treated  as  real  forces  on  the
same  footing  as  any  other  forces  of  nature.  The  reason  for  the  occurrence  in
accelerated systems of reference of such peculiar forces should, according to this new
idea,  be  sought  in  the  circumstance  that  the  distant  masses  of  the  fixed  stars  are
accelerated  relative  to  these  systems  of  reference.  The  \virg{fictitious}  forces  are,  thus,
treated as a kind of gravitational force, the acceleration of the distant masses causing a
\virg{field of gravitation} in the reference frame considered. A recent discussion of the  relativity  of  rotational  motion  within  the  context  of
general relativity can be found in \citet{Gron}. For a detailed historical review of the consequences of rotating masses in relativistic physics, including also the centrifugal-type effects of order $\mathcal{O}(\Psi^2)$, see \citet{Pfister}; here we will briefly review just some salient points. Limiting to the larger first-order effects in rotation $\Psi$, with the so-called\footnote{It is a theory in which the left-hand side of the field equations is a \virg{tensor} which is covariant only with respect to a reduced class of coordinate transformations.} \virg{Entwurf} tensorial theory \citep{entwu}, developed with M. Grossmann in 1913, Einstein and M. Besso   considered a spherical rotating mass shell in  the so-called Einstein-Besso manuscript\footnote{Such an important document has been reprinted and commented in \citet{taccuino}.}, pp. 36-37, and worked out the resulting Coriolis-like dragging force occurring inside. Such a result was publicly presented by \citet{vienna}
in his talk in September 1913 at the Naturforscherversammlung in Vienna\footnote{H. Thirring attended it.}. Einstein, presumably realizing that the Coriolis-type effects in the final version of his theory of gravitation  would not have differed qualitatively from his previous results in the Entwurf theory \citep{Pfister}, did not repeat the calculations after he published his general relativity in November 1915. Such a task was undertaken by H. Thirring in 1917 who, anyway, initially neglected the effects of order $\mathcal{O}(\Psi)$ concentrating on the the much smaller centrifugal ones of order $\mathcal{O}(\Psi^2)$ \citep{taccuino2}. It was after the letter by Einstein of 2 August 1917 \citep{lettera} to him that Thirring was put on the track of the Coriolis-type dragging effects in general relativity. He\footnote{Thirring did not acknowledge the contributions by Einstein to this specific topic \citep{Pfister}.} computed them
in \citet{Thirring} by obtaining a result only differing by a multiplicative factor of two from the Entwurf-based calculation by Einstein: the general relativistic dragging inside a massive rotating hollow  shell is
\eqi \bds A=-2(\bds \Xi\bds\times\bds v),\lb{thir}\eqf
with
\eqi\bds\Xi = q\bds\Psi,\ q\doteq \rp{4G\mathfrak{M}}{3 c^2 R},\lb{rota}\eqf
where $G$ is the Newtonian constant of gravitation, $\mathfrak{M}$ and $R$ are the mass and the radius of the shell, respectively, and $c$ is the speed of light in vacuum; $\Xi$ has the dimensions of s$^{-1}$.
The validity of \rfr{thir} carries over to all regions in which $R\ll r$. The adimensional coefficient $q$ in \rfr{rota} was modified as
\eqi q\doteq \rp{4\varepsilon(2-\varepsilon)}{(1+\varepsilon)(3-\varepsilon)},\eqf with
\eqi\varepsilon\doteq \rp{G\mathfrak{M}}{2c^2 R}, \eqf by \citet{Brill}. It is interesting to note that by choosing a suitably rotating frame with angular velocity $\Psi^{'}$ it is possible to eliminate the Coriolis-force inside the shell. In principle, the results by \citet{vienna} and \citet{Thirring} concerning the rotating hollow shell can be criticized from a Machian point of view because of the asymptotic flatness of the exterior solution, instead of using cosmological boundary conditions. Anyway, in recent times it was demonstrated that, essentially, the Coriolis-like dragging effects inside the mass shell in an asymptotically flat background carry over with minor changes to cosmological boundary conditions. For example, \citet{Klein} showed that it is possible to embed a slowly rotating massive shell with flat interior in a rotationally perturbed Friedmann universe obtaining dragging effects comparable to those by \citet{Thirring}; of course, they depend on the type of the Friedmann universes ($k=0,\pm 1$), and on its mass density.

Putting aside the localized Einstein-Thirring rotating mass shell, in the early years of general relativity attempts were made to construct exact or approximate solutions of the field equations with rotating matter source having cosmological significance; after all, rotation is an ubiquitous phenomenon in nature. We can observe rotating objects at all scales of the Universe,
from the elementary particles to planets, stars and galaxies. The question is, whether this property is an attribute of
the whole universe at a very large scale structure \citep{Whit}. On the other hand, if our universe does not rotate, then it should be explained why and how this happens. Since rotation is generic in the universe, the possible rotation of the universe cannot
be excluded at the very beginning. Moreover, we should explain the physical mechanism which prevents universal
rotation of the universe \citep{Obu02}. \citet{Lanczos} was the first to consider the possibility of a globally rotating universe modeled as a rigidly rotating dust cylinder of infinite radius\footnote{A serious drawback of such a model was that the dust density  diverges at radial infinity.}. For a general overview on physical foundations and observational effects of cosmic rotation, see, e.g., \citet{Obu}. After early speculations concerning the role of cosmic rotation on the galaxy formation \citep{Gamow} and the universe's structure \citep{Wei}, \citet{Goedel}, proposed a stationary cosmological model
  %
  %
  %
  %
in which $\Psi$, the angular velocity of the cosmic rotation, is a positive constant having dimensions of s$^{-1}$.
\citet{Goedel} considered the simplest matter source, i.e. ideal dust with the energy-momentum tensor $T_{\mu\nu}=\rho u_{\mu}u_{\nu}$, and by solving the Einstein field equations with a cosmological constant $\Lambda$ obtained\footnote{For G\"{o}del $\Lambda$ is negative.}
\eqi \rp{\Psi^2}{c^2}=-\Lambda.\eqf Two problems of the solution by \citet{Goedel} were the lack of expansion of the resulting model of the universe, and the violation of causality due to the existence of closed timelike curves. It was later demonstrated that suitable extensions of the original  model, including, e.g., more general energy-momentum tensors or cosmic shear, may circumvent such issues \citep{Obu}.
The cosmic rotation affects  the polarization of the radiation propagating in the curved spacetime giving rise to  an observable effect. Concerning the magnitude of the cosmic vorticity, it turns out to be of the order of the Hubble parameter \citep{hubblo} $H_0=71.0 $ km s$^{-1}$ Mpc$^{-1}=2.3\times 10^{-18}$ s$^{-1}=1.5$ mas cty$^{-1}$. Indeed, \citet{Obu0} yields\footnote{Here $l$ and $b$ denote the Galactic longitude and latitude, respectively.} $\Psi=(1.8\pm 0.8)H_0$, $l=295^{\circ}\pm 25^{\circ}, b=24^{\circ}\pm 20^{\circ}$ on the basis of the data by \citet{Birch}; \citet{Obu}, interpreting the data by \citet{Nod} concerning the dipole effect of the rotation of the plane of polarization as arising from cosmic rotation, finds  $\Psi=(6.5\pm 0.5)H_0$, $l=50^{\circ}\pm 20^{\circ}, b=-30^{\circ}\pm 25^{\circ}$. Within the errors,  the directions of $\bds\Psi$ are orthogonal to each other.
Local, astronomical consequences of the \citet{Goedel} model have been recently investigated in  \citet{Blome} by looking for a possible explanation of the Pioneer anomaly occurring in the remote regions of the solar system.

Recently, investigations of rotational perturbations of pure Friedmann-Lema\^{\i}tre-Robertson-Walker (FLRW) cosmologies, by discarding also in this case the local rotating mass shell, and of their Machian impact have been produced \citep{Ceko,Schmid0,Schmid1,Schmid2}. In the framework of the  FLRW models with global rotation,   \citet{prepolacchi,polacchi0,polacchi} tackled the problem of
the recently observed cosmic acceleration from supernov{\ae} SNIa and Cosmic Microwave Background (CMB). \citet{polacchi} showed that the acceleration of the universe can be explained in terms of the global rotation of the universe.
Moreover, global rotation gives a natural explanation of the empirical relation between angular momentum for clusters and superclusters of galaxies \citep{polacchi}.
Some studies of the CMB
polarization induced by the global rotation have been published \citep{pola}; for earlier investigations on this topic, see \citet{Hawking,Barrow}.
Other cosmological features like the recent discoveries of some non-Gaussian properties of the Cosmic Microwave Background Anisotropies (CMBA), such as the suppression of the quadrupole and the alignment of some multipoles have attracted further attention to rotationally perturbed FLRW models with a cosmological constant \citep{cinesi}.
By comparing the second-order Sachs-Wolfe effect due to rotation with the CMBA data,  \citet{cinesi} constrained the angular speed of the rotation to be less than $20$ mas cty$^{-1}=3.2\times 10^{-17}$ s$^{-1}$ at the last scattering surface.

In all the cases considered so far, a rotation is induced with respect to a locally quasi-inertial Fermi frame;
 thus, a Coriolis-type acceleration\footnote{This is a first-order effect, not to be confused with the second-order ones due to the tidal forces induced in the local frame by the background cosmological space-time; see, e.g., \citet{Cooper}.} affecting the motions of test particles moving with respect to it arises \citep{Blome}. See, e.g., \citet{Silk} for early studies of the Coriolis and centrifugal accelerations acting on a test particle moving in the G\"{o}del spacetime. This fact opens, in principle, interesting perspectives to put local constraints on
the angular velocity vector $\Psi$ in a purely phenomenological way from local dynamics of the planets of the solar system.
Recall that the celestial reference system (ICRS) used is based on a kinematical definition, making the axis directions fixed with respect to the distant matter of the universe. The system is materialized by a celestial reference frame (ICRF) defined by the precise coordinates of extragalctic objects, mostly quasars, BL Lac sources and few active galactic nuclei (AGNs). The current positions are known to better than a milliarcsecond (mas)= $4.8\times 10^{-9}$ rad. According to the IAU recommendations, the origin is to be at the barycenter of the solar system, and the directions of the axes should be fixed with respect to the quasars via VLBI observations.
Such recommendations further stipulate that the principal plane should be as close as possible to the mean equator at J2000.0 and that the origin of it should be as close as possible to the dynamical equinox of J2000.0.
It turns out that the uncertainty from the representation of the ICRS is smaller than 0.01 mas, and the axes are stable to $\pm 0.02$ mas. Note that this frame stability
is based upon the assumption that the extragalctic sources have no proper motion and that there is no global rotation of the universe \citep{IERS}.
It must be noted that the aforementioned accuracies make, in principle, meaningful the analysis proposed; indeed, $\Psi=\sqrt{-\Lambda} c =3.4\times 10^{-18}$ s$^{-1}=2.2$ mas cty$^{-1}$ in view of the currently accepted value of the cosmological constant $-\Lambda=1.26\times 10^{-52}$ m$^{-2}$ \citep{LAMBDA}.

The paper is organized as follows. In Section \ref{dua} we analytically work out the long-term effects on the orbital motion of a test particle perturbed by a small Coriolis-type acceleration. In Section \ref{tria} we compare the resulting predictions with the latest observational determinations of the non-standard perihelion precessions of some planets of the solar system. Section \ref{quatra} is devoted to summarizing our results and to the conclusions.

\section{Analytical calculation}\lb{dua}
The orbital effects induced on a test particle by any acceleration $\bds A$ quite smaller than the dominant Newtonian monopole  $-GM/r^2$ of the primary of mass $M$, whatever its physical origin may be, can be worked out with standard perturbative techniques by using, e.g., the Gauss equations for the variations of the Keplerian orbital elements \citep{Ber}. They are
\begin{eqnarray}\lb{Gauss}
\dert a t & = & \rp{2}{n\eta} \left[e A_R\sin f +A_{T}\left(\rp{p}{r}\right)\right],\lb{gaus_a}\\  \nonumber \\
\dert e t  & = & \rp{\eta}{na}\left\{A_R\sin f + A_{T}\left[\cos f + \rp{1}{e}\left(1 - \rp{r}{a}\right)\right]\right\},\lb{gaus_e}\\ \nonumber \\
\dert I t & = & \rp{1}{na\eta}A_N\left(\rp{r}{a}\right)\cos u,\\  \nonumber \\
\dert\Omega t & = & \rp{1}{na\sin I\eta}A_N\left(\rp{r}{a}\right)\sin u,\lb{gaus_O}\\   \nonumber \\
\dert\omega t & = &\rp{\eta}{nae}\left[-A_R\cos f + A_{T}\left(1+\rp{r}{p}\right)\sin f\right]-\cos I\dert\Omega t,\lb{gaus_o}\\  \nonumber \\
\dert {\mathcal{M}} t & = & n - \rp{2}{na} A_R\left(\rp{r}{a}\right) -\eta\left(\dert\omega t + \cos I \dert\Omega t\right),\lb{gaus_M}
\end{eqnarray}
where $a,e,I,\Omega,\omega,{\mathcal{M}}$ are the semimajor axis, the eccentricity, the inclination, the longitude of the ascending node, the argument of pericenter and the mean anomaly, respectively, of the orbit of the test particle. Moreover,  $f$ is its true anomaly reckoned from the pericentre position, $u\doteq\omega+f$ is the argument of latitude, $n\doteq\sqrt{GM/a^3}=2\pi/P_{\rm b}$ is the unperturbed Keplerian mean motion related to the unperturbed Keplerian orbital period $P_{\rm b}$, $\eta\doteq\sqrt{1-e^2}$ and $p\doteq a(1-e^2)$ is the semi-latus rectum. Finally, $A_R,A_T,A_N$ are the projections of the perturbing acceleration $\bds A$ onto the radial $R$, transverse $T$ and out-of-plane $N$ directions of the particle's co-moving frame $\{\bds{\hat{r}},\bds{\hat{\tau}},\bds{\hat{\nu}}\}$.
Since in the following we will work out the net, secular effects of $\bds A$ on the particle's Keplerian orbital elements, the right-hand sides of \rfr{gaus_a}-\rfr{gaus_M} have to be evaluated onto the unperturbed Keplerian trajectory. To this aim, it will turn out to be convenient to use the eccentric anomaly $E$ instead of the true anomaly $f$;
useful conversion relations are
\begin{eqnarray}
r &=& a(1-e\cos E),\\ \nonumber \\
  \cos f &=& \rp{\cos E-e}{1-e\cos E}, \\ \nonumber \\
  \sin f &=& \rp{\eta \sin E}{1-e\cos E}, \\ \nonumber \\
  dt &=& \rp{(1-e\cos E)}{n}dE.
\end{eqnarray}

In our case, the disturbing acceleration is a phenomenological Coriolis-like one
\eqi\bds A=-2\bds \Psi\bds\times \bds v,\lb{corio}\eqf
where $\Psi$ has the dimensions of s$^{-1}$, and, in general, we do not make any a-priori assumption on the rotation velocity vector $\bds \Psi$, i.e., we pose
\eqi \bds \Psi = \Psi_x\bds i +\Psi_y\bds j+\Psi_z\bds k.\eqf Actually, in some models of rotating universe $\bds\Psi$ depends both on time and radius \citep{cinesi}, but given the typical temporal and spatial scales of the solar system orbital motions, we can safely assume it to be a constant vector.
The components of the planet's velocity $\bds v$ entering \rfr{corio} can be evaluated onto the unperturbed Keplerian ellipse
as
\begin{eqnarray}
  v_x &=& \rp{\partial x}{\partial E}\dert{E}{t}, \\ \nonumber \\
  v_y &=& \rp{\partial y}{\partial E}\dert{E}{t}, \\ \nonumber  \\
  v_z &=& \rp{\partial z}{\partial E}\dert{E}{t},
\end{eqnarray}
in which
\begin{eqnarray}
  x &=& r\left(\cos\Omega\cos u\ -\cos I\sin\Omega\sin u\right),\\ \nonumber \\
 y &=& r\left(\sin\Omega\cos u + \cos I\cos\Omega\sin u\right),\\ \nonumber \\
 z &=& r\sin I\sin u,
\end{eqnarray}
and
\eqi \dert{E}{t}=\rp{n}{1-e\cos E}.\eqf
Note that, since we are going to take an average over one orbital revolution of the right-hand sides of the Gauss equations, we can consider $I,\Omega, \omega$ as constant, so that the partial derivatives of $x,y,z$ with respect to $E$ involve only $r$ and $u$.

The time-varying unit vectors of the co-moving frame along the radial, transverse and normal directions are \citep{Mont}
\eqi \bds{\hat{r}} =\left(
       \begin{array}{c}
          \cos\Omega\cos u\ -\cos I\sin\Omega\sin u\\
          \sin\Omega\cos u + \cos I\cos\Omega\sin u\\
         \sin I\sin u \\
       \end{array}
     \right)
\eqf
 \eqi \bds{\hat{\tau}} =\left(
       \begin{array}{c}
         -\sin u\cos\Omega-\cos I\sin\Omega\cos u \\
         -\sin\Omega\sin u+\cos I\cos\Omega\cos u \\
         \sin I\cos u \\
       \end{array}
     \right)
\eqf
 \eqi \bds{\hat{\nu}} =\left(
       \begin{array}{c}
          \sin I\sin\Omega \\
         -\sin I\cos\Omega \\
         \cos I\\
       \end{array}
     \right)
\eqf
Thus, the projections of the perturbing acceleration on them, defined as
\begin{eqnarray}
  A_R &\doteq & \bds A\bds\cdot \bds{\hat{r}} \\
  A_T &\doteq & \bds A\bds\cdot \bds{\hat{\tau}} \\
  A_N &\doteq & \bds A\bds\cdot \bds{\hat{\nu}},
\end{eqnarray}
are of the form
\eqi A_i =\sum_{j=\{x,y,z\}} \mathcal{V}_{ij}(a,e,I,\Omega,\omega; E) \Psi_j,\ i=R,T,N,\eqf
where the coefficients ${\mathcal{V}}_{ij}$ have dimensions of velocities.
The coefficients of the radial acceleration $A_R$ are
\begin{eqnarray}\lb{CR}
\mathcal{V}_{Rx}&=&\rp{2a^2 n\eta\sin I\sin\Omega}{r},\\ \nonumber \\
\mathcal{V}_{Ry}&=& -\rp{2a^2 n\eta\sin I\cos\Omega}{r},\\ \nonumber \\\
\mathcal{V}_{Rz}&=& \rp{2a^2 n\eta\cos I}{r}.
\end{eqnarray}
The transverse acceleration $A_T$ is built of
\begin{eqnarray}\lb{CT}
\mathcal{V}_{Tx}&=&-\rp{2a^2 e n\sin I\sin\Omega\sin E}{r},\\ \nonumber \\
\mathcal{V}_{Ty}&=& \rp{2a^2 e n\sin I\cos\Omega\sin E}{r},\\ \nonumber \\\
\mathcal{V}_{Tz}&=& -\rp{2a^2 e n\cos I\sin E}{r}.
\end{eqnarray}
The coefficients of the normal acceleration $A_N$ are
\begin{eqnarray}\lb{CN}
%
  \mathcal{V}_{Nx} & = & \rp{2 a^2 n}{r}\left[ \sin E \left( \cos\Omega\sin\omega+\cos I\sin\Omega\cos\omega \right) - \right. \nonumber \\
    & & \left.- \eta\cos E \left( \cos\Omega\cos\omega-\cos I\sin\Omega\sin\omega \right) \right], \\ \nonumber \\
  \mathcal{V}_{Ny} & = & -\rp{2 a^2 n}{r}\left[\sin E\left(\cos I\cos\Omega\cos\omega-\sin\Omega\sin\omega\right)+\right. \nonumber \\
   & & \left. +\eta\cos E\left(\cos I\cos\Omega\sin\omega + \sin\Omega\cos\omega\right)\right] \\ \nonumber \\
  \mathcal{V}_{Nz} & = & -\rp{2a^2 n\sin I}{r}\left(\sin E\cos\omega+\eta\cos E\sin\omega\right)\lb{ultima}.
\end{eqnarray}

In order to make contact with the latest observational determinations from planetary motions, it is convenient to work out the secular precession of the longitude of the pericenter $\varpi$ defined as
\eqi\varpi\doteq \Omega+\omega.\eqf From \rfr{gaus_O}-\rfr{gaus_o} it turns out that, actually, a small perturbing Coriolis-like acceleration induces a non-zero secular precession of $\varpi$ given by
\eqi\left\langle\dot\varpi\right\rangle=\sum_{j=x,y,z}{\mathcal{P}}_j(I,\Omega) \Psi_j,\lb{equz}\eqf
with the adimensional coefficients ${\mathcal{P}}_j$ given by
\begin{eqnarray}
  \mathcal{P}_x &=& -\tan\left(\rp{I}{2}\right)\sin\Omega,\lb{nx} \\  \nonumber \\
   \mathcal{P}_y &=& \tan\left(\rp{I}{2}\right)\cos\Omega, \\ \nonumber \\
   \mathcal{P}_z &=& -1\lb{nz}.
\end{eqnarray}

Also other Keplerian orbital elements, for which no observational investigations yet exist, undergo secular changes.
It turns out that no secular variations occur for the semimajor axis $a$ and the eccentricity $e$.
Instead, the inclination $I$ experiences a secular rate given by
\eqi\left\langle\dot I\right\rangle = \sum_{j=x,y,z}{\mathcal{I}}_j(\Omega) \Psi_j,\lb{noduz}\eqf
with
\begin{eqnarray}
  \mathcal{I}_x &=& -\cos\Omega,\lb{Ix} \\  \nonumber \\
   \mathcal{I}_y &=& -\sin\Omega, \\ \nonumber \\
   \mathcal{I}_z &=& 0\lb{Iz}.
\end{eqnarray}
The node $\Omega$ precesses at a rate
\eqi\left\langle\dot\Omega\right\rangle = \sum_{j=x,y,z}{\mathcal{N}}_j(I,\Omega) \Psi_j,\lb{noduz}\eqf
with
\begin{eqnarray}
  \mathcal{N}_x &=& \cot I\sin\Omega,\lb{nox} \\  \nonumber \\
   \mathcal{N}_y &=& -\cot I\cos\Omega, \\ \nonumber \\
   \mathcal{N}_z &=& -1\lb{noz}.
\end{eqnarray}
The secular change of the mean anomaly $\mathcal{M}$ is
\eqi\left\langle\dot{\mathcal{M}}\right\rangle = n+\sum_{j=x,y,z}{\mathcal{M}}_j(I,\Omega) \Psi_j,\lb{manoz}\eqf
with
\begin{eqnarray}
  \mathcal{M}_x &=& -\eta\csc I\left(\cos I+3\sin^2 I\right)\sin\Omega,\lb{mox} \\  \nonumber \\
   \mathcal{M}_y &=& \eta\left(\cot I+3\sin I\right)\cos\Omega, \\ \nonumber \\
   \mathcal{M}_z &=& \eta\left(1-3\cos I\right)\lb{moz}.
\end{eqnarray}
\section{Confrontation with the latest observational determinations}\lb{tria}
Recently, \citet{Pit010} has analyzed more than 550000 planetary observations of several kinds covering the time interval $1913-2008$. She used the dynamical force models of the EPM2008 ephemerides by estimating about 260 parameters.
The reference frame used to numerically integrate the equations of motion, assumed non-rotating with respect to extra-galactic sources and, thus, locally inertial, is a barycentric one aligned with ICRF by including into the total solution also the VLBI data of spacecraft near the planets; actually, the fundamental plane is the mean ecliptic at J2000.0, so that the $z$ axis points towards the North Ecliptic Pole, in Draco constellation, with right ascension $\alpha=18^{\rm h} 0^{\rm m} 0.0^{\rm s}$ and declination $\delta = +66^{\circ} 33^{'} 38.6^{''}$.
Among the estimated parameters, she also determined corrections $\Delta\dot\varpi$ to the standard Newtonian/Einsteinain perihelion precessions of all the planets of the solar system including Pluto as well; such corrections, by construction, account for any unmodelled/mismodelled dynamical effects, so that they can be used, in principle, to preliminarily put constraints on $\bds \Psi $. Actually, the entire planetary data set should be re-processed with $ad-hoc$ modified models to account for the Coriolis-type effect investigated; one or more dedicated solve-for parameters should be simultaneously estimated along with all the other ones, but this lies outside the scopes of the present paper. In order to constrain $\bds \Psi$, we will use the inner planets, whose estimated perihelion corrections are listed in Table \ref{tavola}, because they are more accurate.
\begin{table}
\caption{Estimated corrections $\Delta\dot\varpi$, in mas cty$^{-1}$, to the standard perihelion precessions with the EPM2008 ephemerides. The quoted errors are not the formal, statistical ones but are realistic. From Table 8 of \citet{Pit010}.}
\label{tavola}
\begin{tabular}{llll}\hline
Mercury & Venus & Earth & Mars\\
\hline
$-4\pm 5$  & $24\pm 33$  & $6\pm 7$ & $-7\pm 7$\\
\hline
\end{tabular}
\end{table}
In Table \ref{tavola2} we quote the coefficients $\mathcal{P}_j$ of the Coriolis-type perihelion precessions for the inner planets whose relevant Keplerian orbital elements are listed in Table \ref{tavola3}.
\begin{table}
\caption{Longitude of the ascending node $\Omega$ and inclination $I$, both in deg, for the inner planets. Reference frame: ICRF/J2000. Coordinate system: Ecliptic and Mean Equinox of Reference Epoch. From Table A.2 of \citet{Mur}.}
\label{tavola3}
\begin{tabular}{lll}\hline
& $\Omega$ & $I$  \\
\hline
Mercury & $48.33167$ & $7.00487$ \\
Venus & $76.68069$ & $3.39471$ \\
Earth &  $348.73936$ & $0.00005$\\
Mars & $49.57854$ & $1.85061$\\
\hline
\end{tabular}
\end{table}
\begin{table}
\caption{Coefficients $\mathcal{P}_j, \ j=x,y,z$ for the inner planets computed from Table \ref{tavola3} and \rfr{nx}-\rfr{nz}.}
\label{tavola2}
\begin{tabular}{llll}\hline
& $\mathcal{P}_x$ & $\mathcal{P}_y$ & $\mathcal{P}_z$ \\
\hline
Mercury & $-0.04572069$  & $0.04069034$  & $-1$ \\
Venus & $-0.02883601$ & $0.00682681$ & $-1$\\
Earth & $8\times 10^{-8}$ & $4.3\times 10^{-7}$ & $-1$\\
Mars & $-0.01229570$ & $0.01047240$ & $-1$\\
\hline
\end{tabular}
\end{table}

By equating the corrections $\Delta\dot\varpi$ of Table \ref{tavola} to the predicted Coriolis-type precessions $\left\langle\dot\varpi\right\rangle$ of \rfr{equz} for three planets it is possible to set up a non-homogeneous algebraic linear system in the three unknowns $\Psi_x,\Psi_y,\Psi_z$.
By using Mercury, Venus and the Earth allows to obtain
\begin{eqnarray}
  \Psi_x &=& -0.9\pm 1.8\ {\rm arcsec\ cty}^{-1}, \lb{psix1}\\ \nonumber \\
  \Psi_y &=& -1.3\pm 2.0\ {\rm arcsec\ cty}^{-1}, \\ \nonumber \\
  \Psi_z &=& -0.006\pm 0.007\ {\rm arcsec\ cty}^{-1},\lb{psiz1}
\end{eqnarray}
with
\eqi \Psi = 1.6\pm\ 2.6\ {\rm arcsec\ cty}^{-1}.\lb{merda}\eqf
The quoted uncertainties have been obtained by linearly propagating the errors in $\Delta\dot\varpi$ of Table \ref{tavola}.
It turns out that the other combinations involving Mercury, Venus and Mars, and Venus, Earth and Mars
yield similar constraints, while those from
Mercury, Earth and Mars are about one order of magnitude weaker.
The tightest constrain holds for $|\Psi_z|\leq 0.013$ arcsec cty$^{-1}$; instead, $\Psi_x,\Psi_y$ may be as large as $\approx 1$ arcsec cty$^{-1}$. Recall that
1 arcsec cty$^{-1}$ = $1.5\times 10^{-15}$ s$^{-1}$.

In Table \ref{tavolaold} we quote the values of the corrections $\Delta\dot\varpi$ estimated by \citet{Pit05} with older versions of the EPM ephemerides and less extended data sets; for Venus, Earth and Mars they are, for some reasons, more accurate by about one order of magnitude than the more recent results of Table \ref{tavola}.
\begin{table}
\caption{Corrections $\Delta\dot\varpi$, in milliarcsec cty$^{-1}$, to the standard perihelion precessions estimated by E.V. Pitjeva with the EPM2005 (Mercury, Earth, Mars) and EPM2006 (Venus) ephemerides. The quoted errors are not the formal, statistical ones but are realistic. From Table 3 of \citet{Pit05} (Mercury, Earth, Mars) and Table 4 of \citet{Fienga} (Venus).}
\label{tavolaold}
\begin{tabular}{llll}\hline
Mercury & Venus & Earth & Mars\\
\hline
$-3.6\pm 5.0$  & $-0.4\pm 0.5$  & $-0.2\pm 0.4$ & $0.1\pm 0.5$\\
\hline
\end{tabular}
\end{table}
The data of Table \ref{tavolaold}, applied to Mercury, Venus, Earth, yield
\begin{eqnarray}
  \Psi_x &=& -0.02\pm 0.08\ {\rm arcsec\ cty}^{-1}, \lb{psix2}\\ \nonumber \\
  \Psi_y &=& -0.1\pm 0.2\ {\rm arcsec\ cty}^{-1}, \\ \nonumber \\
  \Psi_z &=& -0.0002\pm 0.0004\ {\rm arcsec\ cty}^{-1}\lb{psiz2},
\end{eqnarray}
corresponding to
\eqi \Psi = 0.1\pm\ 0.2\ {\rm arcsec\ cty}^{-1},\lb{merdaccia}\eqf
of the order of $10^{-16}$ s$^{-1}$, i.e.
one order of magnitude more accurate that the limit from \rfr{merda}.

By comparison, \citet{Folk} phenomenologically tested if the locally inertial reference frame in which the orbits of the planets are usually integrated is not rotating with
respect to the rest of the universe. He
compared the mean motion of Mars relative to Earth determined from ranging measurements
with the mean motion determined from VLBI observations of Mars-orbiting
spacecraft relative to stars or extra-galactic radio
sources. \citet{Folk} finds that the dynamical rotation
rate of the solar system relative to extra-galactic radio sources is less than 0.004 arcsec cty$^{-1}=6\times 10^{-18}$ s$^{-1}$. Actually, more details on the approach followed by \citet{Folk} would be needed to make a meaningful comparison with our results. In fact, no long-term effects may occur for the mean motion because the Coriolis force does not secularly affect the semimajor axis $a$; instead, the mean longitude $\ell\doteq \varpi+\mathcal{M}$ experiences a secular variation. A result quantitatively similar to that by \citet{Folk} could be obtained with our method by a-priori setting $\Psi_x=\Psi_y=0$ because, in this case, it would simply be
\begin{eqnarray}
  \dot\varpi &=& -\Psi, \\ \nonumber \\
  \dot \ell &\approx & -3\cos I \Psi.
\end{eqnarray}

Interpreted in terms of rotation of the universe \citep{Blome}, our results show that local astronomical effects in the solar system do not show statistically significant evidence for its existence, putting independent upper bounds on its magnitude. They are not in contrast with, e.g., the G\"{o}del-type expected value  $\Psi=2$ mas cty$^{-1}$ and with the upper bound of 0.02 arcsec cty$^{-1}$ by \citet{cinesi}.
\citet{Obu0} and  \citet{Obu} yield not only the magnitude of the cosmic rotation but also its spatial direction. More precisely,
from the Galactic longitude $l$ and latitude $b$ of $\bds\Psi$ released by \citet{Obu0} it is possible to infer its direction with respect to the solar system barycentric frame used
\begin{eqnarray}
  \hat{\Psi}_x &=& -0.73\pm 0.24, \\ \nonumber \\
  \hat{\Psi}_y &=& -0.51\pm 0.34, \\ \nonumber \\
  \hat{\Psi}_z &=& -0.45\pm 0.32,
\end{eqnarray}
so that, by summing in quadrature the errors in $\Psi$ and in the components of the unit vector,
\begin{eqnarray}
  \Psi_x &=& -1.97\pm 1.09\ {\rm mas\ cty}^{-1}\lb{obu0x}, \\ \nonumber \\
  \Psi_y &=& -1.38\pm 1.10\ {\rm mas\ cty}^{-1} , \\ \nonumber \\
  \Psi_z &=& -1.21\pm 1.02\ {\rm mas\ cty}^{-1}.\lb{obu0z}
\end{eqnarray}
It can be noted that \rfr{obu0x}-\rfr{obu0z} are compatible with both \rfr{psix1}-\rfr{psiz1} and the more stringent bounds of \rfr{psix2}-\rfr{psiz2}.
The values by \citet{Obu} for $l$ and $b$
yield for the components of the unit vector of $\bds\Psi$
\begin{eqnarray}
  \hat{\Psi}_x &=& 0.72\pm 0.27, \\ \nonumber \\
  \hat{\Psi}_y &=& -0.63\pm 0.30, \\ \nonumber \\
  \hat{\Psi}_z &=& 0.27\pm 0.34.
\end{eqnarray}
Thus,
\begin{eqnarray}
  \Psi_x &=& 7.0\pm 2.7\ {\rm mas\ cty}^{-1}\lb{obux}, \\ \nonumber \\
  \Psi_y &=& -6.1\pm 2.9\ {\rm mas\ cty}^{-1} , \\ \nonumber \\
  \Psi_z &=& 2.6\pm 3.3\ {\rm mas\ cty}^{-1}.\lb{obuz}
\end{eqnarray}
Also in this case, \rfr{obux}-\rfr{obuz} are compatible with both \rfr{psix1}-\rfr{psiz1} and \rfr{psix2}-\rfr{psiz2}.
It is remarkable to note that our approach, in the framework of the polarization radiation due to cosmic rotation, is able to put local, solar system-scale constraints on the Hubble parameter $H_0$ which are just $2-3$ orders of magnitude larger that its value determined from cosmological observations.
  Incidentally, let us note that the relation between the G\"{o}delian universe's vorticity and the cosmological constant allows us to put local constraints on it much tighter than those usually obtained from the Schwarzschild-de Sitter metric usually adopted so far. Indeed, from \rfr{merda} it follows
\eqi-\Lambda \leq 2\times 10^{-46}\ {\rm m}^{-2},\eqf
while \rfr{merdaccia} yields
\eqi-\Lambda \leq 1\times 10^{-48}\ {\rm m}^{-2};\eqf
on the contrary, the solar system-based constraints obtained from the perturbations by the Hooke-type radial acceleration due to the Schwarzschild-de Sitter metric are of the order of\footnote{Here $\Lambda$ is defined positive.} \citep{Iorio06,jez,kag,ser}
\eqi\Lambda \leq 10^{-40}-10^{-42}\ {\rm m}^{-2}.\eqf
With regard to the connection with cosmological parameters, interesting perspectives may open up in near-mid future if and when the interplanetary laser ranging technique \citep{PLR} will be implemented allowing for a notable improvement of the planets'orbit determination.

Concerning the application of our Coriolis-type results to the rotation of the solar system through the Milky Way \citep{Blome}, let us start by noting that it occurs clockwise about the North Ecliptic Pole, \citep{galass}, located in the constellation Coma Berenices. Thus, the associated unit vector\footnote{Conventionally, it is directed towards the South Galactic Pole, located in the constellation Sculptor, so that the Sun's rotation appears anticlockwise from its tip.} $\bds{\hat{\Psi}}$ has ecliptic longitude and latitude
\begin{eqnarray}
  \lambda_{\bds{\hat{\Psi}}} &=& 0.02319\ {\rm deg}, \\ \nonumber \\
  \beta_{\bds{\hat{\Psi}}} &=& -29.81149\ {\rm deg}
\end{eqnarray}
and components
\begin{eqnarray}
  \hat{\Psi}_x &=& 0.867, \\ \nonumber \\
  \hat{\Psi}_y &=& 3\times 10^{-4}, \\ \nonumber \\
  \hat{\Psi}_z &=& -0.497.
\end{eqnarray}
The magnitude of the Galactic angular velocity can  approximately\footnote{A circular motion is assumed.} be evaluated as $v_{\odot}/(2\pi r_{\odot})$, so that
\eqi \Psi= (1.54\pm 0.19)\times 10^{-16}\ {\rm s}^{-1}=0.101\pm 0.009\ {\rm arcsec\ cty}^{-1},\lb{omegaz}\eqf
where we used $v_{\odot}=(254\pm 16)$ km s$^{-1}$, $r_{\odot}=(8.4\pm 0.6)$ kpc \citep{gala}.
Its components are
\begin{eqnarray}
  \Psi_x &=& 0.087\pm 0.008\ {\rm arcsec\ cty}^{-1},\lb{galx}\\ \nonumber \\
  \Psi_y &=& \mathcal{O}(10^{-5})\ {\rm arcsec\ cty}^{-1},\\ \nonumber \\
  \Psi_z &=& -0.050\pm 0.004 \ {\rm arcsec\ cty}^{-1}\lb{galz}.
\end{eqnarray}
It must be noted that \rfr{galz} is statistically incompatible with  both \rfr{psiz1} and \rfr{psiz2} at more than $3-\sigma$ level.

It may be interesting to point out that if we interpret our results in terms of an Einstein-Thirring rotating massive shell identified with the Oort cloud \citep{Opik, Oort}, we get unphysical results. Indeed, by assuming \citep{Weiss} $\mathfrak{M}_{\rm Oort}\approx 38 m_{\oplus}$ and $R\approx 10^4$ au, \rfr{rota}, yields
\eqi q_{\rm Oort}\approx 1.5\times 10^{-16}.\eqf Thus, \rfr{merda} and \rfr{merdaccia} would yield an angular velocity of the Oort shell as large as $1-10$ s$^{-1}$. Of course, it is very daring to identify an extended object of certainly non-uniform density like the Oort cloud with the Thirring's infinitely thin
massive shell of uniform density.

\section{Summary and conclusions}\lb{quatra}
We analytically worked out the secular precessions of the Keplerian orbital elements of a test particle affected by a small extra-acceleration of Coriolis type treated perturbatively with the Gauss variational equations. It could be due to a relative rotation of the reference frame considered with respect to distant masses; we reviewed several theoretical local and cosmological scenarios leading to such an effect. In the calculation we did not make any a-priori assumptions on the spatial orientation of the angular velocity vector $\bds\Psi$ which was, thus, treated as a constant vector directed along a generic direction in space.
We found that the semimajor axis $a$ and the eccentricity $e$ do not undergo secular changes, while the inclination $I$, the longitude of the ascending node $\Omega$, the longitude of the pericenter $\varpi$ and the mean anomaly $\mathcal{M}$ experience secular variations which are linear combinations the components of $\bds\Psi$ with coefficients depending on $e,I,\Omega$.

Then, we compared the Coriolis-induced theoretical prediction of the precession of $\varpi$ with the latest observational determinations of the corrections $\Delta\dot\varpi$ to the standard Newtonian-Einsteinian precessions of the inner planets of the solar system estimated by fitting long data sets with different versions of the EPM ephemerides. No statistically significative evidence for a non-zero angular velocity $\bds\Psi$ was found; thus, we were able to constrain  $\bds\Psi$. The tightest bounds occur for $\Psi_z$ whose magnitude can be as large as
$0.0006-0.013$ arcsec cty$^{-1}$. The other components can be as large as $|\Psi_x|\leq 0.1-2.7$ arcsec cty$^{-1}$ and $|\Psi_y|\leq 0.3-2.3$ arcsec cty$^{-1}$. It must be pointed out that such upper bounds should be considered just as order-of-magnitude figures; indeed, the entire planetary data set should be re-processed by explicitly modeling the effect we are interested in, and dedicated solve-for parameters should be simultaneously estimated in a new global solution along with all the other ones routinely determined.

It turns out that several values of the cosmic rotation obtained from cosmological observations interpreted in terms of different theoretical models are compatible with our results.
In the framework of the rotating G\"{o}del universe, they yield local, astronomical constraints on the cosmological constant $\Lambda$ of the order of $10^{-46}-10^{-48}$ m$^{-2}$, i.e.  several orders of magnitude tighter than those usually obtained so far from the Schwarzschild-de Sitter metric. On the other hand, interpreting our results in terms of the polarization of radiation propagating in a rotating universe allows to put solar system-scale constraints on the Hubble parameter $H_0$ which are just $2-3$ orders of magnitude larger that its value derived from cosmological observations. This opens interesting perspectives in view of the expected future improvements in the planetary orbit determination from the implementation of the interplanetary laser ranging technique. Concerning the effect of the rotation of the solar system through the Milky Way, occurring clockwise about the North Galactic Pole, it turns out that the corresponding component along the $z$ axis of $\Psi$ is in disagreement with our values for $\Psi_z$ at more than $3-\sigma$ level.
Finally, let us mention that the application of our investigations to the Oort cloud, modeled as an Einstein-Thirring rotating massive shell inducing Coriolis-type forces inside, yields  unphysical values for its putative rotation.

\section*{Acknowledgments}
I thank an anonymous reviewer for useful critical remarks.


\end{document}